\pdfoutput=1

\documentclass[%
aps,%
prb,%
twocolumn,%
footinbib,%
bibnotes,%
showpacs,%
floatfix]%
{revtex4}
\usepackage{graphicx}
\usepackage{stmaryrd}
\usepackage{color}%
\usepackage{amsmath,amssymb}
\usepackage{hyperref}

\definecolor{darkred}{rgb}{0.5,0.0,0.0}
\definecolor{darkgreen}{rgb}{0.0,0.5,0.0}
\definecolor{darkblue}{rgb}{0.0,0.0,0.5}

\hypersetup{colorlinks=true,
citecolor=darkred,
linkcolor=darkgreen,
urlcolor=darkblue}

\newcommand{\tmfloatsmall}[2]{
\begin{figure}
#1
\caption{(Color online) #2}
\end{figure}}

\newcommand{\tmfloatsmallfirst}[2]{
\begin{figure}[b]
#1
\caption{(Color online) #2}
\end{figure}}

\newcommand{\emdash}{---}

\newcommand{\mathi}{\mathrm{i}}

\newcommand{\tmop}[1]{\ensuremath{\operatorname{#1}}}

\begin{document}

\title{Modeling extended contacts for nanotube and graphene
devices}

\author{Norbert Nemec}
\affiliation{TCM Group,
 Cavendish Laboratory,
 19 JJ Thomson Avenue,
 Cambridge CB3 0HE, UK.}
\affiliation{Institute for Theoretical Physics,
 University of Regensburg,
 D-93040 Regensburg, Germany}

\author{David Tom\'anek}
\affiliation{Physics and Astronomy Department,
 Michigan State University,
 East Lansing, MI 48824-2320, USA.}
\affiliation{Institute for Theoretical Physics,
 University of Regensburg,
 D-93040 Regensburg, Germany}

\author{Gianaurelio Cuniberti}
\affiliation{Max Bergmann Center for Biomaterials,
 Dresden University of Technology,
 D-01062 Dresden, Germany.}
\affiliation{Institute for Theoretical Physics,
 University of Regensburg,
 D-93040 Regensburg, Germany}

\date{\today}

\begin{abstract}
  Carrier injection into carbon nanotubes and graphene nanoribbons, contacted
  by a metal coating over an arbitrary length, is studied by various means:
  Minimal models allow for exact analytic solutions which can be transferred
  to the original system with high precision. Microscopic \textit {ab initio}
  calculations of the electronic structure at the carbon-metal interface allow
  us to extract---for Ti and Pd as contacting materials---realistic
  parameters, which are then used in large scale tight-binding models for
  transport calculations. The results are shown to be robust against
  nonepitaxially grown electrodes and general disorder at the interface, as
  well as various refinements of the model.
\end{abstract}


\maketitle

\section{Introduction}

The high electrical conductivity of metallic carbon nanotubes (CNTs) can be
attained, thanks to a unique combination of several features. The
quasi-one-dimensional crystal structure, together with a low density of
defects, allows us to explore the theoretical limit of conductance of $4 e^2 /
h$ at the charge neutrality point. The stiffness of carbon-carbon bonds
reduces the effect of electron-phonon coupling at room temperature
.\cite{gheorghe-veitlcocn2005} Also, restricting electron movement to a
single dimension results in a very small phase space, which strongly reduces
the effectiveness of scattering. A further reduction of backscattering is
caused by the low density of states at the Fermi energy in combination with a
high Fermi velocity. Considering all these factors, measured ballistic lengths
of several microns\cite{mann-btimnwrpoc2003} become understandable. Yet, to
exploit the potential for carrying current densities of up to $10^9 ~
\mathrm{A} / \mathrm{\tmop{cm}}^2$,\cite{yao-hetiscn2000} the contacts at a
nanometer scale become crucial.

For graphene, which has been shown to allow ballistic transport over similar
lengths,\cite{geim-trog2007} the contacts play an equally important
role.\cite{schomerus-ecmfttwg2007} Finite-width graphene nanoribbons (GNRs)
are discussed as nanoelectronic devices and interconnects, showing both
similarities and distinctive differences when compared to nanotubes. With
first experimental results becoming available,\cite{chen-gne2007} many of
the theoretical predicions about GNRs will soon be put to
test.\cite{nakada-esigrnseaesd1996,fujita-plsazge1996,son-hgn2006,wimmer-stirgn2008}

Experimentally, a crucial factor for obtaining good metallic contacts are the
wetting properties of the material. Thus, it has been observed that Ti, Ni and
Pd form continuous coatings on single-wall CNTs while Au, Al and Fe form
isolated particles.\cite{zhang-mcoscnaiitmi2000} Furthermore, among several
common contact metals, Ti was found to be the only one where true chemical
bonds could be observed, while the others showed only weak van der Waals
interactions.\cite{zhang-fomnosscn2000} Surprisingly enough,
Pd{\emdash}traditionally known as a rather poor conductor{\emdash}was found to
form better and more reliable Ohmic contacts for CNTs than
Ti\cite{mann-btimnwrpoc2003} and could be successfully applied to produce a
CNT field effect transistor with Ohmic
contacts.\cite{javey-cnftwiocahgd2004} It is generally believed that this
superiority of Pd is due to its high work function ($\phi_{\tmop{Pd}} = 5.1 ~
\mathrm{\tmop{eV}}$) that matches well with that of CNTs [e.g.,
$\phi_{\text{$(7, 0)$~CNT}} = 5.1 ~ \mathrm{\tmop{eV}}$
(Ref.~{\onlinecite{shan-fpsowfoswcn2005}})] and thereby avoids a high
Schottky barrier. Pt, which has an even higher work function, would therefore
be expected to perform even better as a contacting material, but as it turns
out it does not form Ohmic contacts at all.\cite{javey-bcnft2003}

\tmfloatsmallfirst{\resizebox{3.375in}{!}{\includegraphics{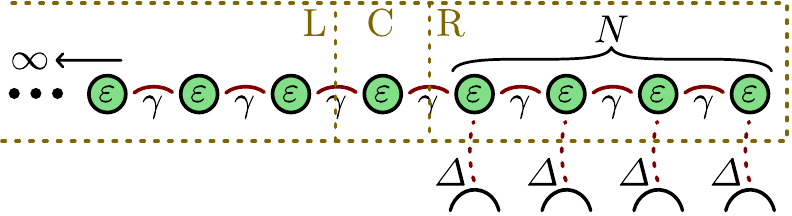}}}{\label{fig:linchain-scheme}Minimal
model for extended contacts solvable analytically: a linear chain of identical
atoms (one orbital per atom) with hopping integral $\gamma$ between nearest
neighbors. To the left, the chain continues infinitely; at the right end, $N$
atoms are contacted, each by an independent wideband lead of strength
$\Delta$. For defining the transmission, the system is virtually split into
three regions: the conductor $C$ and the leads $L$ and $R$.}

A further experimental puzzle is the question of the effective length of
contacts formed by a coating metal layer: While some studies report that
transport occurs only at the edge of the contact,\cite{mann-btimnwrpoc2003}
others state that the contact resistance depends on the length of the
contact\cite{wakaya-cromcn2003} or that extended contacts show a clearly
distinct behavior from pointlike contacts.\cite{chiu-cnnits2007}

Various theoretical studies have been conducted to investigate these issues:
In \textit {ab initio} studies comparing Au, Pd and Pt contacting an $(8,
0)$~CNT, Pd was found to have the lowest Schottky
barrier.\cite{shan-aisosbamc2004} Another \textit {ab initio} study
comparing the metal-graphene bonding of the same three metals indicated a very
small binding energy for Au. For Pd it is somewhat larger, while for Pt the
bonding is yet stronger.\cite{maiti-mibeawp2004} The bad contacts formed by
Pt are here attributed to a clustering effect of larger metal grains. Direct
\textit {ab initio} simulations of transport in a metal-contacted CNT
compared Pd and Au,\cite{ke-nj2a3tet2006,palacios-mcicnftbtsbp2008} finding
again that Pd forms superior contacts. Ti was also found to form strong bonds
to a nanotube surface\cite{meng-fsocbtsascn2007} and to form superior point
contacts for CNTs.\cite{liu-aisotscn2003}

Apart from these practical issues in explaining and improving the quality of
materials, the study of contact models is also of great theoretical relevance:
In studying the physics of electronic devices at the nanometer scale, it is
generally crucial to have detailed control over the
contacts.\cite{hipps-me-aac2001,cuniberti-trocime2002} Indeed, specifying
properties of nanoelectronic devices is generally completely meaningless
without clearly stating the way the system was contacted or{\emdash}for
theoretical studies{\emdash}how the contact was modeled.

In a previous work,\cite{nemec-cdociicnaais2006} we have demonstrated the
counterintuitive finding that the optimal transparency in extended contacts
for CNTs is achieved not by strong chemical bonding---as it would be the case
for pointlike
contacts\cite{liu-aisotscn2003,krompiewski-eocioqcoan2003,chibotaru-ettacn2003,deretzis-rocboetimns2006}---but
rather by contact materials that gently couple to the tube surface, allowing
us to exploit the length of the tube-metal interface to smoothly inject the
electrons with minimal reflection at the contact.

In this paper, we will present a detailed analysis of the model used for our
previous findings. A minimal model, reducing the CNT or GNR to a single atomic
chain and considering a semi-infinite wire with only a single contact at one
end, allows an analytic solution that gives detailed insight in the mechanism
of an extended contact (see Fig.~\ref{fig:linchain-scheme}). The results from
this model have a close relation to the physics of Breit-Wigner resonances,
which will be briefly sketched out at the beginning. After a thorough analysis
of the minimal model, we will demonstrate how the results can be transferred
to more realistic structures and explain in detail how the missing parameters
could be quantiatively estimated from \textit {ab initio} calculations of Pd
and Ti as two typical contacting materials.

\section{Breit-Wigner resonance}

The conductance for a molecular junction, consisting of a single energy level
$\varepsilon$ in a two-terminal setup between two leads, is given in terms of
the left and right tunneling rates $\Gamma_{\mathrm{L}}$ and
$\Gamma_{\mathrm{R}}$ by the Breit-Wigner
equation\cite{breit-cosn1936,stone-eoiportiod1985,garca-caldern-doorimts1993}
\begin{eqnarray}
  G \left( E \right) & = & \frac{2 e^2}{h} \frac{\Gamma_{\mathrm{L}}
  \Gamma_{\mathrm{R}}}{\left( E - \varepsilon \right)^2 + \left(
  \Gamma_{\mathrm{L}} + \Gamma_{\mathrm{R}} \right)^2 / 4} . 
  \label{eq:Breit-Wigner}
\end{eqnarray}
The bell-shaped peak in this expression as a function of the energy is well
known. What is rarely noted in literature, however, is the fact that also for
fixed energy $E$ and one fixed contact $\Gamma_{\mathrm{L}}$, the conductance
as a function of the other contact $G \left( \Gamma_{\mathrm{R}} \right)$ has
a bell shape with an optimum at the balanced coupling $\Gamma_{\mathrm{R}} =
\Gamma_{\mathrm{L}}$. A new perspective to this old issue was provided in the
recent experiments by Gr\"uter et al.\cite{grter-rttacmjiale2005} Small
couplings ($\Gamma_{\mathrm{R}} \ll \Gamma_{\mathrm{L}}$) result in a linear
$\Gamma_{\mathrm{R}}$ dependence of the conductance typical of tunneling
phenomena, while for large coupling ($\Gamma_{\mathrm{R}} \gg
\Gamma_{\mathrm{L}}$), such better contact $\Gamma_{\mathrm{R}}$ results in an
overall suppressed conductance. One way to understand this counterintuitive
behavior is to consider the tunneling rate $\Gamma$ as a measure for the
chemical bond between the conducting orbitals of the molecule and the lead: A
strong bond to one of the leads causes the molecule itself to virtually become
part of that lead so that we observe the physics of a single point contact.
Furthermore, the strong bonding results in a strong redistribution of the
spectral weight of the energy level in the molecule, i.e., in a low local
density of states (LDOS) at the energy $\varepsilon$. The tunneling
conductance, which directly probes this LDOS, will therefore be suppressed by
large $\Gamma_{\mathrm{R}}$.

Once the length $N$ of a contact is increased for an extended molecule, the
optimal value decreases monotonically with the number of contact points $N$,
as displayed in Fig.~\ref{fig:breit-wigner-scheme}. As will be shown later
[see Eq.~(\ref{Delta-opt})] this value scales like $\Gamma_{\mathrm{R}} =
\Gamma_{\mathrm{L}} \ln N / N$ for large $N$.

\tmfloatsmall{\resizebox{3.375in}{!}{\includegraphics{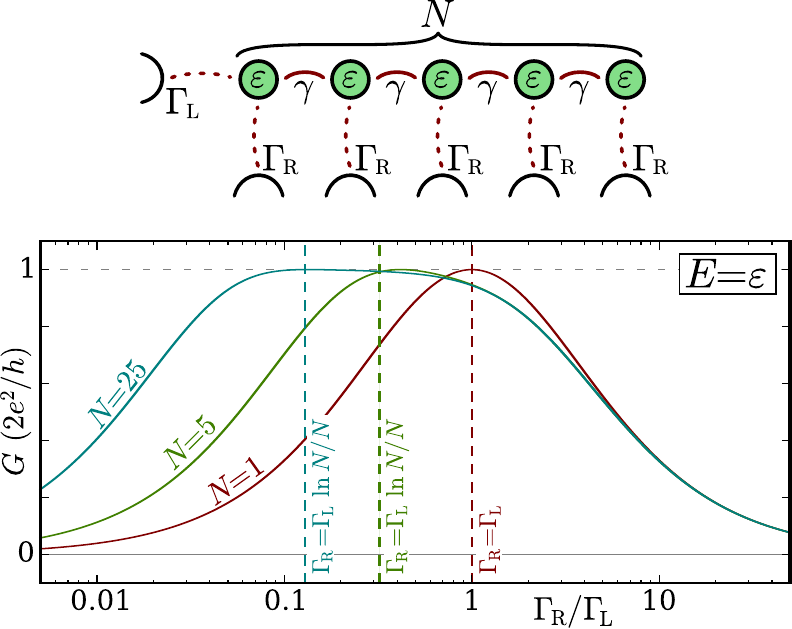}}}{\label{fig:breit-wigner-scheme}Breit-Wigner
resonance in the conductance of an extended molecule sandwiched between two
leads. The internal hopping is fixed as $2 \gamma = \Gamma_{\mathrm{L}}$ to
allow an optimal match to the left contact. In the case $N = 1$, the system is
identical to the molecular junction described by Eq.~(\ref{eq:Breit-Wigner}).
The transmission shows the shift of the Breit-Wigner peak towards lower
$\Gamma_{\mathrm{R}}$ with growing $N$. The functional form of this shift can
be approximated for large $N$ as $\Gamma_{\mathrm{R}} = \Gamma_{\mathrm{R}}
\ln N / N$ [see Eq.~(\ref{Delta-opt})].}

\section{Analytical model}

A minimal model that captures the essential physics of extended contacts is
set up as follows. The CNT or GNR is represented by a linear chain of atoms
with the hopping integral $\gamma$ and the onsite energy $\varepsilon = 0$
(fixing the energy offset). A two-probe setup is defined by selecting an
arbitrary single electron as the ``conductor'' and the semi-infinite sections
at both ends as ``leads''. In this unmodified setup, the system is fully
transparent, so the transmission $T \left( E \right)$ is equal to the number
of channels $N_{\tmop{ch}}$ at any given energy. The single cosine-shaped band
of the linear chain provides a single transmission channel
\begin{eqnarray*}
  T_{\tmop{band}} \left( E \right) & = & \Theta \left( E + 2 \gamma \right)
  \Theta \left( - E - 2 \gamma \right)
\end{eqnarray*}
which presents a theoretical upper transmission limit when scattering at the
contacts could be neglected.

\tmfloatsmall{\resizebox{3.375in}{!}{\includegraphics{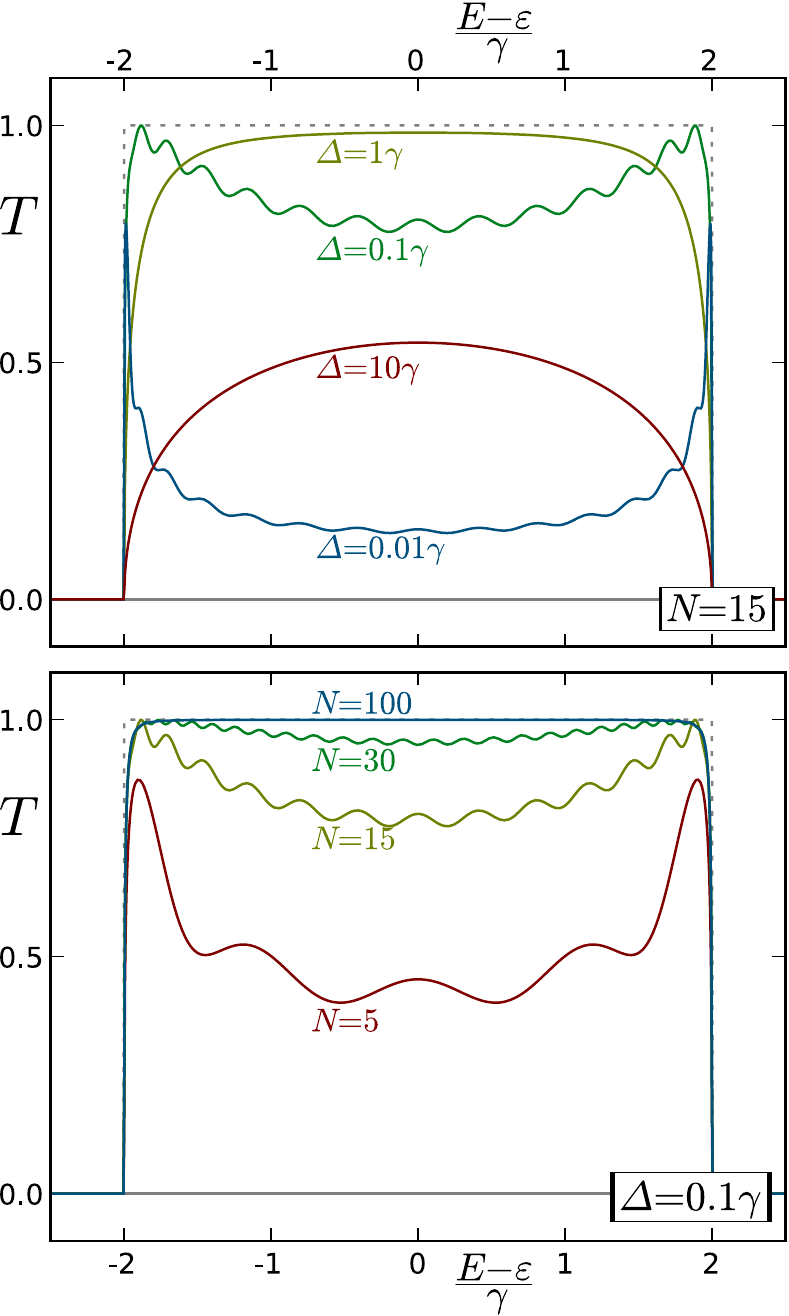}}}{\label{fig:scanE-transmission}Transmission
through the system displayed in Fig.~\ref{fig:linchain-scheme}, as given in
Eq.~(\ref{transmission-chain-to-CWBL}). Top panel: for a fixed contact length
$N$, starting from low $\Delta$, the transmission first improves, reaches an
optimum, and then degrades again at high $\Delta$. Bottom panel: For fixed
contact strength $\Delta$, transmission improves with growing $N$ and
saturates for large $N$.}

An ``extended contact'' to the linear chain is now modeled by replacing the
semi-infinite lead by a finite $N$-atom chain contacted in each atom
individually by a wideband lead of strength $\Delta$. (For a sketch of the
model, see Fig.~\ref{fig:linchain-scheme}.) A full solution of this model is
obtained by calculating the conductance as the quantum mechanical transmission
probability,
\begin{eqnarray*}
  G \left( E \right) & = & \frac{2 e^2}{h} T \left( E \right) .
\end{eqnarray*}
This can be done within the Landauer approach to transport by means of the
Green function formalism, as shown in the Appendix. As a result, one obtains
(for $\varepsilon = 0)$
\begin{eqnarray*}
  T \left( E \right) & = & \frac{8 \sqrt{4 - E^2 / \gamma^2} \tmop{Im} \left[
  f_N \left( E / 2 \gamma - \mathi \Delta / 4 \gamma \right) \right]}{\left| E
  / \gamma - \mathi \sqrt{4 - E^2 / \gamma^2} - 2 f_N \left( E / 2 \gamma -
  \mathi \Delta / 4 \gamma \right) \right|^2},
\end{eqnarray*}
with $f_N \left( x \right) = U_{N - 1} \left( x \right) / U_N \left( x
\right)$. $U_N (x)$ are the Chebyshev polynomials of the second kind, as given
in Eq.~(\ref{eqn:chebyshev-recursive}). To gain a full understanding of the
physics described by this expression, the transmission $T (E)$ is plotted as a
function of the energy for different values of the two parameters $N$ and
$\Delta$ (i.e., the length and the quality of the contact region) in
Fig.~\ref{fig:scanE-transmission}. Two regimes can be identified: An
\textit {$N$-resonant} regime for low $\Delta$/small $N$, where the
transmission shows about as many peaks as there are atoms in the contact
region, and an \textit {$N$-independent} regime for high $\Delta$/large $N$,
where the transmission shows no resonances and depends only on $\Delta$.

The two different plots in Fig.~\ref{fig:scanE-transmission} illustrate two
aspects: For fixed $N$ with increasing $\Delta$, the transparency of the
system improves, goes through an optimum point, and degrades again, while for
fixed $\Delta$, the transparency improves with growing $N$ and saturates at an
$N$-independent optimum. In both cases, the transmission goes through the two
different regimes.

\tmfloatsmall{\resizebox{3.375in}{!}{\includegraphics{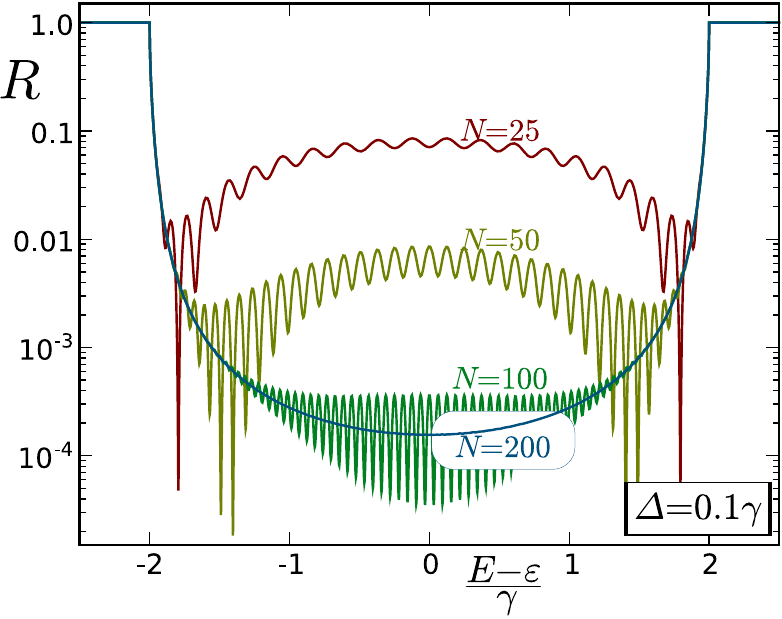}}}{\label{fig:scanE-reflection}Contact
reflection in the system displayed in Fig.~\ref{fig:linchain-scheme}, as given
in Eq.~(\ref{reflection-chain-to-CWBL}). At fixed contact strength $\Delta$,
with growing $N$, the contact becomes more transparent and saturates at an
$N$-independent value.}

Especially the last point can be seen more clearly by looking at the
reflection $R = 1 - T$ in the energy range of the single channel of our
system,
\begin{eqnarray}
  R & = & \left| \frac{E + \mathi \sqrt{4 \gamma^2 - E^2} - 2 \gamma f_N
  \left( E / 2 \gamma - \mathi \Delta / 4 \gamma \right)}{E - \mathi \sqrt{4
  \gamma^2 - E^2} - 2 \gamma f_N \left( E / 2 \gamma - \mathi \Delta / 4
  \gamma \right)} \right|^2  \label{reflection-chain-to-CWBL}
\end{eqnarray}
In Fig.~\ref{fig:scanE-reflection}, this observable is plotted in a
logarithmic scale, illustrating that the average value of the transmission
already saturates at $N = 100$ (specifically for $\Delta = 0.1 \gamma$). For
larger values of $N$, the overall transparency is not improved any further,
but the $N$-dependent resonances are smoothed out.

\tmfloatsmall{\resizebox{3.375in}{!}{\includegraphics{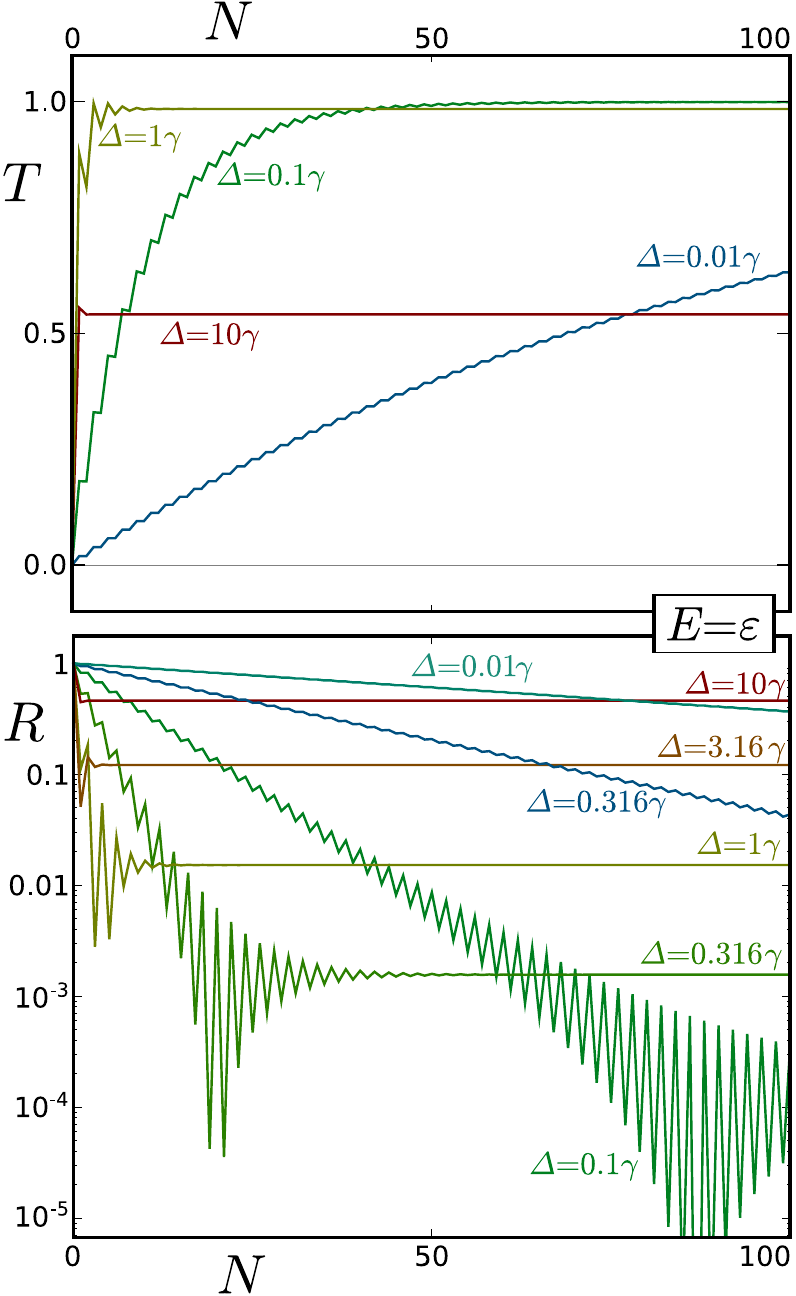}}}{\label{fig:scanN}Transmission
$T$ (top panel) and contact reflection $R$ (bottom panel) in the system
displayed in Fig.~\ref{fig:linchain-scheme} at fixed energy $E = \varepsilon$
for varying contact length $N$ and selected values of the contact strength
$\Delta$.}

To better understand the origin of this saturation, we fix the energy to the
half filling case $E = \varepsilon$ and study the transmission and the
reflection for varying contact lengths $N$ (see Fig.~\ref{fig:scanN}):
\begin{eqnarray}
  R \left( E = \varepsilon \right) & = & \left| \frac{1 + \mathi f_N \left( -
  \mathi \Delta / 4 \gamma \right)}{1 - \mathi f_N \left( - \mathi \Delta / 4
  \gamma \right)} \right|^2 .  \label{R-at-E-eq-epsilon}
\end{eqnarray}
Similar data were numerically obtained before for
CNTs.\cite{nakanishi-cbcname2000} A different insight, however, can be
gained from the reflection in a logarithmic scale: Ignoring the even-odd
oscillations in $N$, one observes first an exponential decay of $R$ with
increasing $N$, followed by an abrupt crossover to an $N$-independent value.
Both the rate of decay and the saturation value depend on $\Delta$ in such a
way that for lower values of $\Delta$, the transparency initially improves
more slowly with the contact length, but ultimately $R$ saturates at a lower
value, which means higher contact transparency. This result, which we
presented before based on numerical calculations on
CNTs,\cite{nemec-cdociicnaais2006} will be studied in more detail in the
following using our analytical expressions.

\tmfloatsmall{\resizebox{3.375in}{!}{\includegraphics{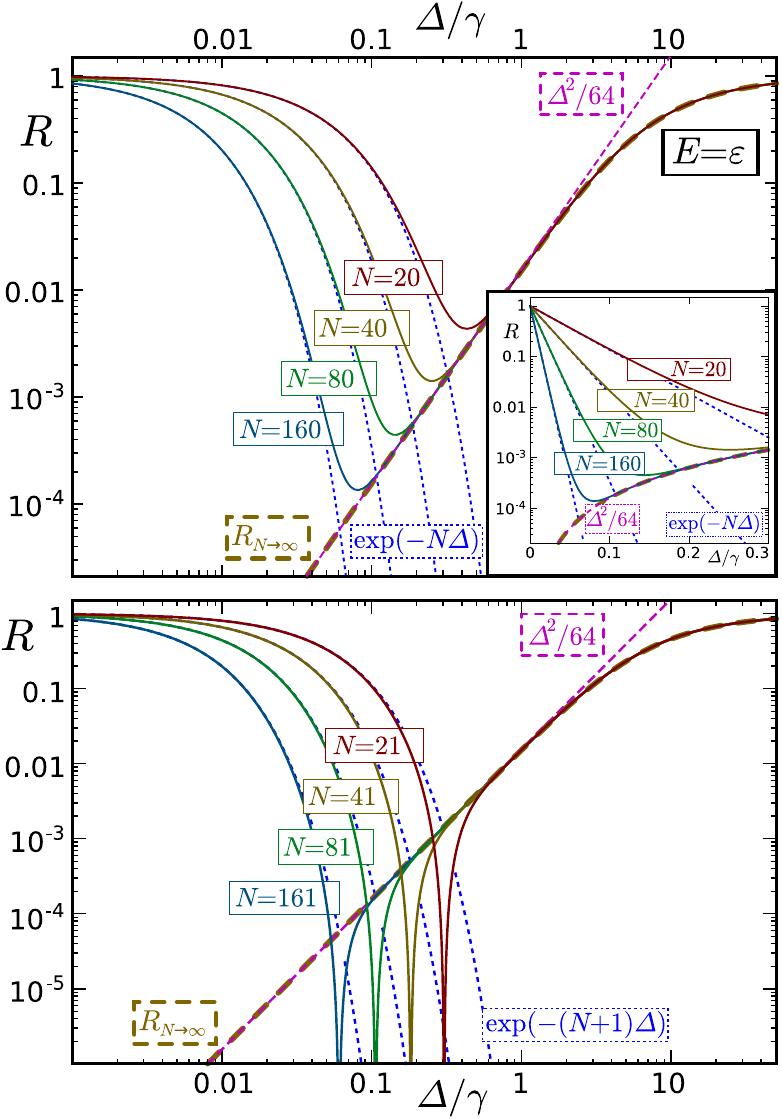}}}{\label{fig:scanDelta}$\Delta$
dependence of the contact reflection $R$ for various even (upper panel) and
odd (lower panel) contact lengths $N$. Solid: the exact value as given in
Eq.~(\ref{reflection-chain-to-CWBL}). Dashed: the limit $R_{N \rightarrow
\infty}$ given in Eq.~(\ref{eqn:R-N-infty}), along with its approximation
$\Delta^2 / 64 \gamma^2$, valid for $\Delta \ll 1$. Dotted: the approximation
$\exp \left( - N \Delta / \gamma \right)$, valid for even $N$ in the
$N$-resonant regime. The inset shows the identical data in the semilogarithmic
scale, further illustrating the precision of the $\exp \left( - N \Delta /
\gamma \right)$ approximation in the $N$-resonant regime.}

An expression for the $N$-independent regime can easily be obtained as the
limit $N \rightarrow \infty$ of Eq.~(\ref{f-infty}) as
\begin{eqnarray}
  R_{N \rightarrow \infty}^{E = \varepsilon} & = & \left| \frac{1 + \mathi
  f_{\infty} \left( - \mathi \Delta / 4 \gamma \right)}{1 - \mathi f_{\infty}
  \left( - \mathi \Delta / 4 \gamma \right)} \right|^2 \nonumber\\
  & = & \left( \frac{\sqrt{\Delta^2 / 4 + 4 \gamma^2} - \Delta / 2 - 2
  \gamma}{\sqrt{\Delta^2 / 4 + 4 \gamma^2} - \Delta / 2 + 2 \gamma} \right)^2,
  \label{eqn:R-N-infty}
\end{eqnarray}
which can be further simplified for $\Delta \ll \gamma$ to obtain
\begin{eqnarray*}
  R^{E = \varepsilon}_{N \rightarrow \infty, \Delta \ll \gamma} & = &
  \frac{\Delta^2}{64 \gamma^2} .
\end{eqnarray*}
The validity of this approximation is illustrated in Fig.~\ref{fig:scanDelta}.

The approach for finding the corresponding approximation for the $N$-resonant
regime is less rigorous since a simple limit is not sufficient to capture the
behavior in this case. A far better approximation is found graphically: The
straight section in a semilogarithmic scale plot (inset of
Fig.~\ref{fig:scanDelta}) indicates a clean exponential law. The missing
coefficients are easily found from a Taylor expansion in $\Delta = 0$,
yielding
\begin{eqnarray*}
  R_{\text{resonant}}^{\text{$N$ even}} & = & \exp \left( - N \Delta / \gamma
  \right),\\
  R_{\text{resonant}}^{\text{$N$ odd}} & = & \exp \left[ - \left( N + 1
  \right) \Delta / \gamma \right], \text{-}
\end{eqnarray*}
both of which can be seen to fit precisely over the whole $N$-resonant region.

Having found good approximations for both regimes, the last missing piece is
the crossover. For even $N$, the smooth shape of the crossover in
Fig.~\ref{fig:scanDelta} suggests a simple function of the form $R =
\sqrt[n]{A^n + B^n}$ and, indeed, we find that for the case $n = 1 / 2$,
\begin{eqnarray}
  R_{\text{crossover}}^{\text{$N$ even}} & = & \left( \sqrt{R_{N \rightarrow
  \infty}} + \sqrt{R_{\text{resonant}}^{\text{$N$ even}}} \right)^2 
  \label{eqn:R-even-crossover}
\end{eqnarray}
gives an extremely good match over the full range of $\Delta$. Moreover, a
very similar function is found to match the crossover for odd values of $N$,
\begin{eqnarray*}
  R_{\text{crossover}}^{\text{$N$ odd}} & = & \left( \sqrt{R_{N \rightarrow
  \infty}} - \sqrt{R_{\text{resonant}}^{\text{$N$ odd}}} \right)^2 .
\end{eqnarray*}
Both approximations show slight deviations from the exact value for small $N$
but match with high precision for larger $N$. Obviously, the two reflection
probabilities behave like squares of quantum mechanical amplitudes interfering
either constructively or destructively with each other.

The parameter values where $R_{\tmop{resonant}}$ and $R_{N \rightarrow
\infty}$ coincide are of special interest. In the case that $N \gg 1$, where
this coincidence happens for $\Delta \ll \gamma$, the condition for this is
simply
\begin{eqnarray*}
  \exp \left( - N \Delta / \gamma \right) & = & \Delta^2 / 64 \gamma^2,
\end{eqnarray*}
leading to an expression for the $\Delta$-dependent effective contact length
\begin{eqnarray}
  N_{\tmop{eff}} \left( \Delta \right) & = & \frac{2 \gamma}{\Delta} \ln
  \left( \frac{8 \gamma}{\Delta} \right),  \label{Neff-of-Delta}
\end{eqnarray}
over which a longer contact does not further modify transport. This can be
interpreted as the length that contributes to the electron transmission for a
very long contact.

The inverse of Eq.~(\ref{Neff-of-Delta}) can expressed using the Lambert-$W$
function,\cite{corless-otlwf1996}
\begin{eqnarray*}
  \Delta_{\tmop{opt}} \left( N \right) & = & 2 \gamma W \left( 4 N \right) / N
\end{eqnarray*}
which can be approximated in the range of interest as
\begin{eqnarray}
  \Delta_{\tmop{opt}} \left( N \right) & \approx & 2 \gamma \ln N / N . 
  \label{Delta-opt}
\end{eqnarray}

\subsection{Generalization to arbitrary injection energies}

Having found the transport relations at the fixed energy $E = \varepsilon$, we
can now continue with generalizing the results for $E \neq \varepsilon$.
Assuming the general functional form of the reflection
\begin{eqnarray*}
  R_{\tmop{resonant}} & = & \exp \left( - 2 N \Delta / \alpha_1 \right)\\
  R_{N \rightarrow \infty} & = & \Delta^2 / \alpha_2^2,
\end{eqnarray*}
we numerically find the following:
\begin{eqnarray*}
  \alpha_1 & = & \sqrt{4 \gamma^2 - E^2} \left( 1 + \zeta \right),
\end{eqnarray*}
which holds for arbitrary fixed $N$ with an approximate error estimate $\left|
\zeta \right| \lesssim 1 / N$ capturing the resonant oscillations. Considering
the characteristic form of the density of states of the linear chain,
\begin{eqnarray*}
  \rho \left( E \right) & = & \left( \pi \sqrt{4 \gamma^2 - E^2} \right)^{-
  1},
\end{eqnarray*}
we can rewrite the last expression as
\begin{eqnarray*}
  \alpha_1 & \approx & 1 / \pi \rho,
\end{eqnarray*}
reflecting the similarity to a weak point contact where the tunneling
transmission is proportional to the density of states on either side. The last
relation could be confirmed numerically to hold very generally, as will be
discussed below in the discussion of realistic contacts for nanotubes and
ribbons.

For the regime of $N \rightarrow \infty$, we can similarly, i.e. numerically,
find the expression
\begin{eqnarray*}
  \alpha_2 & = & \frac{2}{\gamma} \left( 4 \gamma^2 - E^2 \right),
\end{eqnarray*}
which fits the exact formula [Eq.~(\ref{reflection-chain-to-CWBL})] with
arbitrary precision for fixed $\Delta \ll \gamma$ small enough and for $E$ not
too near the band edges. Unlike the formula for $\alpha_1$, however,
expressing $\alpha_2$ as a function of the density of states alone does not
help to generalize the relation to other structures.

The crossover region, generally governed by interference effects, can be
approximated by averaging over quantum mechanical phases, resulting in
\begin{eqnarray*}
  R_{\tmop{crossover}} & = & R_{N \rightarrow \infty} + R_{\tmop{resonant}},
\end{eqnarray*}
which gives a good approximation for the full parameter space with $N \gg 1$,
$\Delta \ll \gamma$ and $E$ away from band edges. Apart from the resonant
oscillations, this now allows the full description of the reflection, and we
find a precise numerical confirmation of the previously obtained expression of
the effective contact length,
\begin{eqnarray*}
  N_{\tmop{eff}} \left( \Delta \right) & = & \frac{1}{\pi \rho \Delta} \ln
  \frac{8 \gamma^2 - 2 E^2}{\gamma \Delta} .
\end{eqnarray*}
A complete overview of the contact reflection and both parameters $N$ and
$\Delta$ is shown in Fig.~\ref{fig:coating-plot2d}.

\tmfloatsmall{\resizebox{3.375in}{!}{\includegraphics{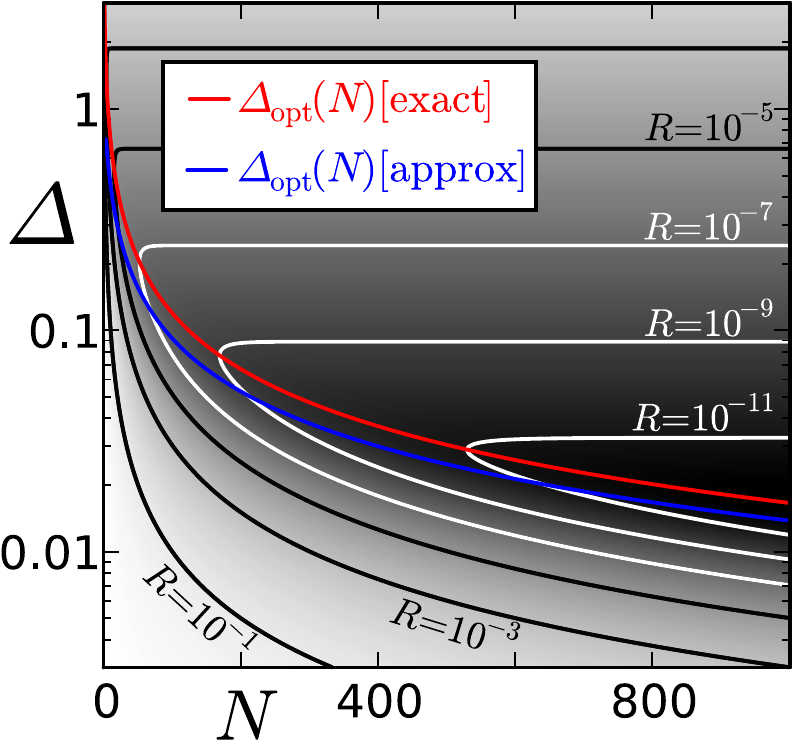}}}{\label{fig:coating-plot2d}The
contact reflection for even $N$ obtained from
Eq.~(\ref{eqn:R-even-crossover}). Quite visible are the two regimes separated
by the minimal line $\Delta_{\tmop{opt}} (N)$. The ``exact'' value for
$\Delta_{\tmop{opt}}$ is the true minimum for fixed $N$. The ``approximate''
value comes from Eq.~(\ref{Delta-opt}).}

\section{Nondiagonal contacts}

\tmfloatsmall{\resizebox{3.375in}{!}{\includegraphics{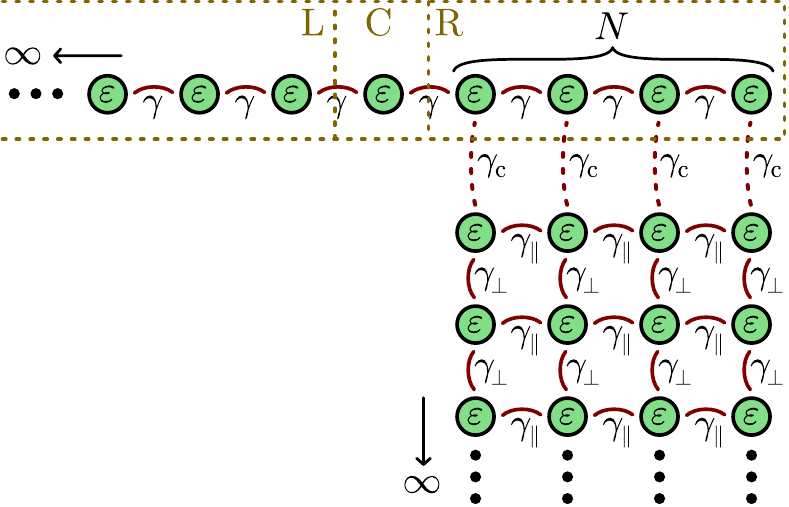}}}{\label{fig:coating-nondiag-scheme}Generalized
model including nondiagonal terms: the individual wideband leads for each atom
in the extended contact region are replaced by a metal with an internal
structure, here modeled as a 2D-square lattice. The new parameters are
$\gamma_{\parallel}$ and $\gamma_{\perp}$, describing the internal hopping in
the lattice parallel and perpendicular to the contact surface, as well as
$\gamma_{\mathrm{c}}$, describing the hopping at the contact. For simplicity,
we consider only the isotropic case $\gamma_{\parallel} = \gamma_{\perp}$.
This leaves us with the single effective parameter $\Delta =
\gamma_{\mathrm{c}}^2 / \gamma_{\perp}$.}

\tmfloatsmall{\resizebox{3.375in}{!}{\includegraphics{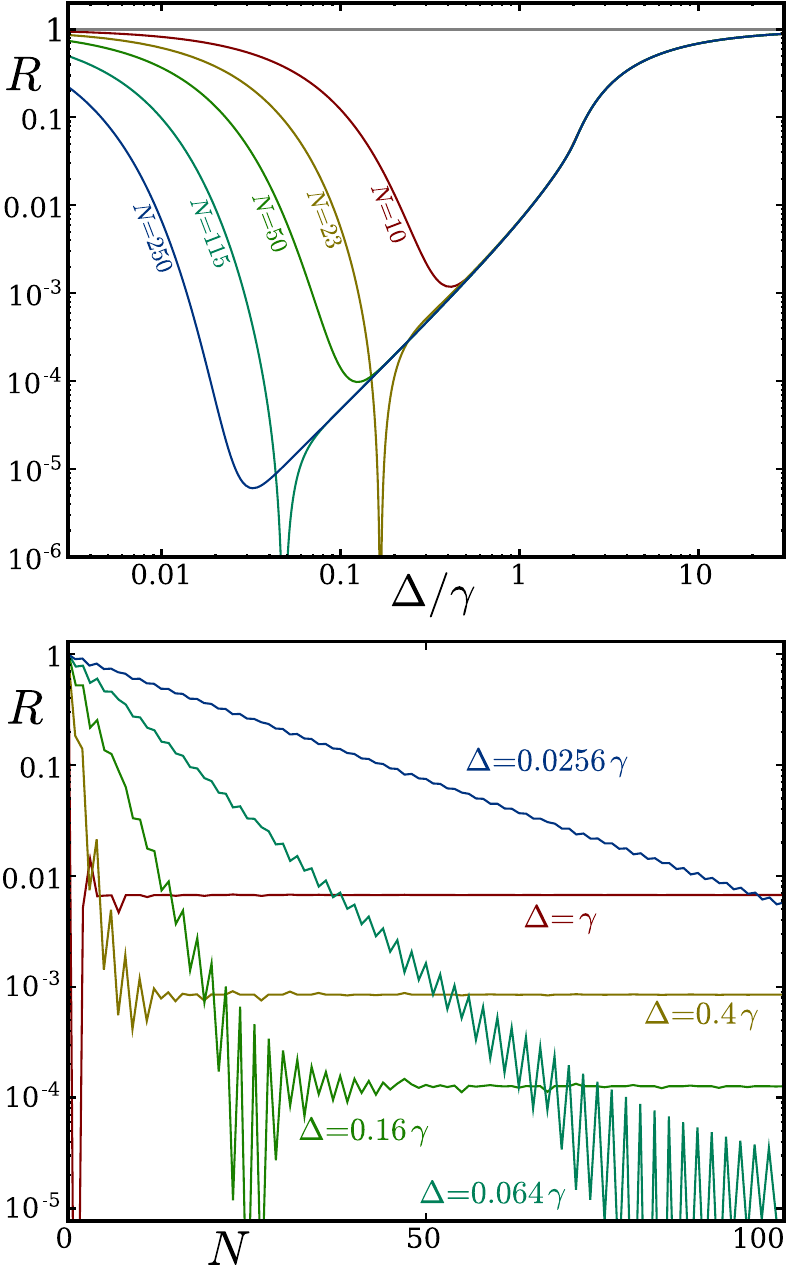}}}{\label{fig:scan-nondiag}Nondiagonal
contacts: Reflection $R$ of the generalized model displayed in
Fig.~\ref{fig:coating-nondiag-scheme}. Relating the parameter $\Delta =
\gamma_c^2 / \gamma_{\perp}$ to the parameter $\Delta$ of the wideband leads,
the results are qualitatively similar to those of the original model
(Figs.~\ref{fig:scanN} and \ref{fig:scanDelta}). One prominent difference is
the enlarged reflection $R (\Delta)$ for $\Delta / \gamma$ between $1$ and
$10$: While the diagonal self-energy was uniform for every atom along the
contact, the nondiagonal self-energy now is sensitive to the edge of the
contact. For large values of $\Delta$, where only the atoms near the edge
contribute to the transport, this causes the visible deviation from the $R =
\Delta^2 / \alpha_2^2$ law. The same reason is behind the visible
irregularities in the resonant oscillations of $R (N)$.}

To generalize our results beyond the diagonal contact approximation, we model
the contacting metal not as a single-parameter wideband lead but as a material
with an internal structure, leading to off-diagonal terms in the contact
matrix (see Fig.~\ref{fig:coating-nondiag-scheme}).
Figure~\ref{fig:scan-nondiag} illustrates that the off-diagonal terms in the
self-energy do not bring any qualitative changes to the behavior described
before. An exact quantitative mapping would depend strongly on the details of
the model.

\tmfloatsmall{\resizebox{3.375in}{!}{\includegraphics{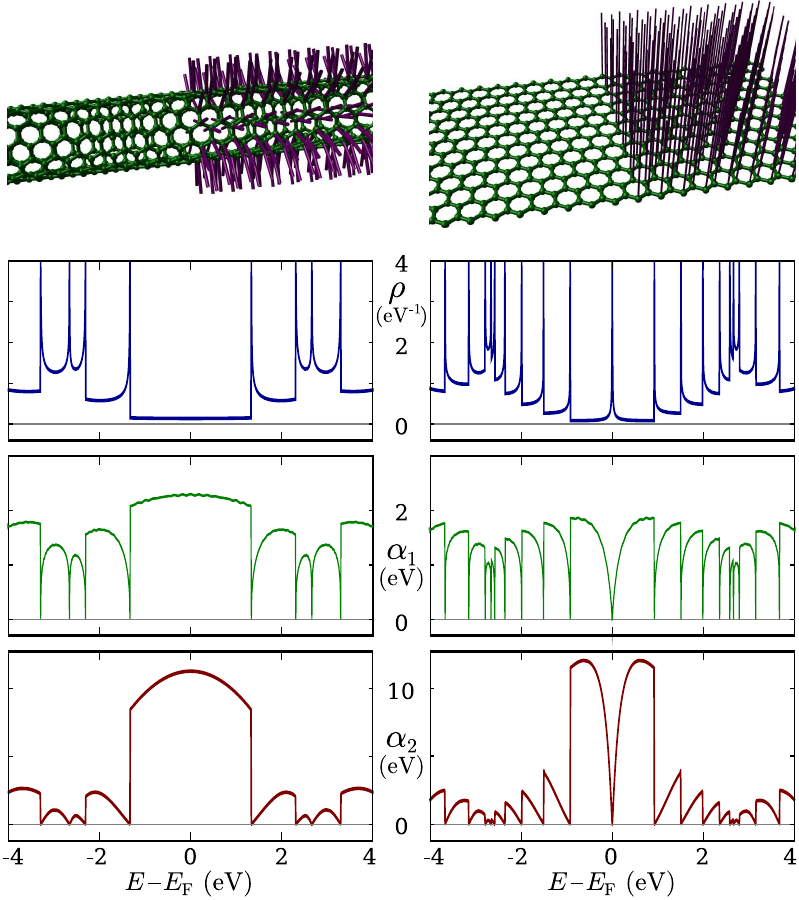}}}{\label{fig:cnt-gnr}Extended
contacts for a (6,6)~CNT (left) and the corresponding graphene nanoribbon. Top
panel: density of states with characteristic van Hove singularities. The
zigzag-edge state in the nanoribbon causes a peak at $E = E_{\mathrm{F}}$.
Center panel: The value $\alpha_1 = - 2 N \Delta / \ln R$, here computed for
$N = 40$ and $\Delta = 10^{- 5} \tmop{eV}$, lies already very near to the
limiting case $\alpha_1 = N_{\tmop{ch}} / \pi \rho$. Bottom panel: The value
$\alpha_2 = \Delta / \sqrt{R}$ is well converged for $N \rightarrow \infty$
and $\Delta = 10^{- 2} \tmop{eV}$. Note the suppression of both $\alpha_1$ and
$\alpha_2$ in the ribbon at $E_{\mathrm{F}}$ where the presence of the
localized edge state suppresses the conductance in the contact region.}

\section{Realistic contacts for carbon nanostructures}

For the case of CNTs and GNRs, the method of the Chebyshev polynomials cannot
be used to obtain an analytical solution due to the the non$\pi
\rho$commutativity of the partial Hamiltonians of the periodic structure. In
numerical studies, however, we find that the behavior is identical to that of
the linear chain, except for a quantitative adjustment of the parameters
$\alpha_1$ and $\alpha_2$ (see Fig.~\ref{fig:cnt-gnr}). The $N$-resonant
regime can be described precisely by a simple generalization of the law found
for the linear chain
\begin{eqnarray*}
  R_{\tmop{resonant}} & = & \exp \left( - 2 N \Delta / \alpha_1 \right),\\
  \alpha_1 & = & N_{\tmop{ch}} / \pi \rho,
\end{eqnarray*}
where $N_{\tmop{ch}}$ is the number of channels and $\rho$ the total density
of states per unit cell. Generally, both values are dependent on the energy
and the chirality of the tube or width and edge geometry of the ribbon. For
metallic CNTs near the Fermi energy, however, one finds the general values of
$N_{\tmop{ch}} = 2$, $\rho = 2 N_{\tmop{ch}} / 3 \gamma d_{\tmop{CC}}$, and,
therefore, $\alpha_1 = 3 \gamma d_{\tmop{CC}} / 2 \pi \ell_{\tmop{uc}}$.
($\gamma = 2.66 ~ \tmop{eV}$ and $d_{\tmop{CC}} = 1.42 ~ \text{{\AA}}$).
Introducing the physical length of the contact region $L = \ell_{\tmop{uc}} N$
with the length of the unit cell $\ell_{\tmop{uc}}$, the previous formula can
be rewritten as
\begin{eqnarray*}
  R_{\tmop{resonant}} & = & \exp \left( - 2 L \Delta / \alpha_1
  \ell_{\tmop{uc}} \right),\\
  \alpha_1 \ell_{\tmop{uc}} & = & 1.80 ~ \tmop{eV} ~ \text{{\AA}} .
\end{eqnarray*}
For the $N$-independent regime, the general law of $R_{N \rightarrow \infty} =
\Delta / \alpha_2$ still holds, but the functional form of the parameter
$\alpha_2$ at arbitrary energies could not be determined. Generally, it turns
out that $\alpha_2$ is suppressed at van Hove singularities in a similar way
as $\alpha_1$ is. Furthermore, metallic CNTs have a fairly constant value of
$\alpha_2$ around $E_F$. At $E = E_{\mathrm{F}}$, we find $\alpha_2 = 4.24
\gamma$ for armchair CNTs \ and $\alpha_2 = 5.66 \gamma$ for metallic zigzag
CNTs.

For GNRs, the situation is slightly more complex due to the presence of edge
states at zigzag
edges.\cite{fujita-plsazge1996,nakada-esigrnseaesd1996,wimmer-stirgn2008}
In metallic ribbons with armchair edges, the situation is similar to that of
metallic CNTs, and we find a value of $\alpha_2 = 8.0 \gamma$ at $E =
E_{\mathrm{F}}$. The quantitative difference from the value of the
corresponding CNTs can be explained by the presence of only one conduction
channel at the Fermi energy. For ribbons with zigzag edges, however, our
simplified model suggests that the constant $\alpha_2$ is completely
suppressed at $E = E_{\mathrm{F}}$ due to the peak in the density of states,
caused by the edge state.

Physically, this suppressed value of both $\alpha_1$ and $\alpha_2$ suggests
that the injection of charges into the edge state via extended leads is less
efficient than that into the conduction channels of nanotubes or armchair
ribbons. However, taking into account the results of detailed \textit {ab
initio} calculations reveals a spin splitting of the edge states, which
strongly affects the bands at the Fermi energy and opens special spin
transport channels,\cite{wimmer-stirgn2008} which are not captured by our
model.

\tmfloatsmall{\resizebox{3.375in}{!}{\includegraphics{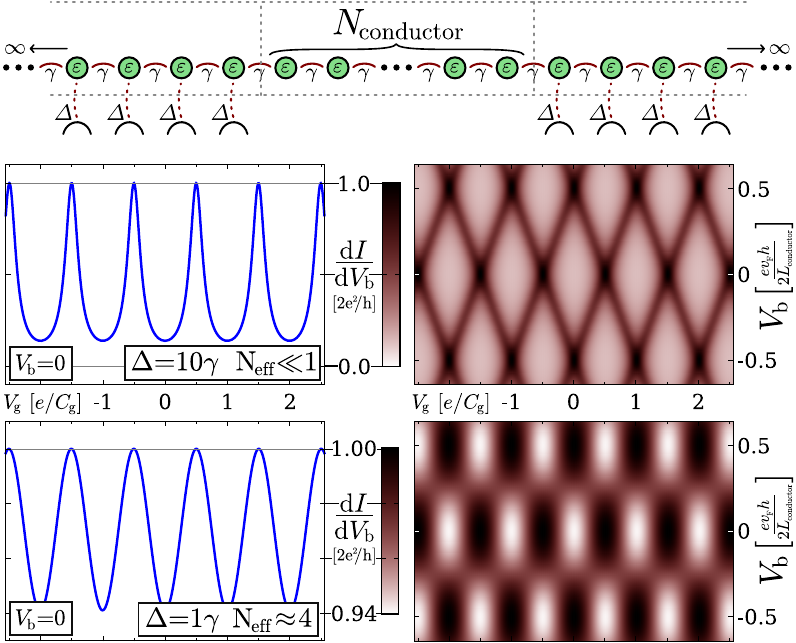}}}{\label{fig:fabry-perot}Differential
conductance through a two-terminal setup of a linear chain with symmetric,
infinite-length extended contacts. The central region consists of
$N_{\tmop{conductor}} = 200$ atoms. The distance between the Fabry-P\'erot
oscillations in the gate voltage $V_{\mathrm{g}}$ depends on the total gate
capacitance $C_{\mathrm{g}}$ as $\delta V_{\mathrm{g}} = e / C_{\mathrm{g}}$.
The extent of the diamonds in direction of the bias voltage $V_{\mathrm{b}}$
depends directly on the level spacing $\delta E = v_{\mathrm{F}} h / 2
L_{\tmop{conductor}} = 2 \gamma \pi / N_{\tmop{conductor}}$ as $\delta
V_{\mathrm{b}} = e \delta E$. Top: strong coupling $\Delta = 10 \gamma$,
leading to an extremely short effective contact length producing sharp
resonances and a distinct diamond pattern. Bottom: moderate coupling $\Delta =
1 \gamma$ leading to an effective contact length of about four unit cells. The
oscillations have a sinoidal shape with strongly reduced amplitude. For yet
weaker coupling as it is to be expected for Pd or Ti, the amplitude is rapidly
reduced even further making the oscillations undetectable. In this case, a
stong defect within the contact region may act as a point of scattering and
recover sharp resonances.}

\section{Two-terminal setup and Fabry-P\'erot physics}

A realistic setup for conduction measurements in CNTs and GNRs generally needs
a second contact at the other end of the system to close a circuit. Such a
setup is well known to lead to Fabry-P\'erot-like oscillations of the
conductance along the energy range.\cite{liang-fiianew2001} For very bad
contacts, Coulomb blockade has been observed, but we intentionally avoid this
regime that would demand the inclusion of charging effects.

One important aspect of Fabry-P\'erot oscillations is their experimental use
in measuring the length of the scattering region. In the zero-bias
differential conductance, the spacing of the Fabry-P\'erot resonances depends
on the gate capacitance $C_{\mathrm{g}}$ alone as $\delta V_{\mathrm{g}} = e /
C_{\mathrm{g}}$. Only the diamond shapes in a plot of the finite bias
differential conductance
\begin{eqnarray*}
  \frac{\mathrm{d} I}{\mathrm{d} V_{\mathrm{b}}} & = & \frac{e^2}{h} \left[ T \left(
  \frac{V_{\mathrm{g}} C_{\mathrm{L}}}{e \rho} - \frac{eV_{\mathrm{b}}}{2}
  \right) + T \left( \frac{V_{\mathrm{g}} C_{\mathrm{L}}}{e \rho} +
  \frac{eV_{\mathrm{b}}}{2} \right) \right]
\end{eqnarray*}
give access to the spacing of the energy levels $\delta E = \hbar
v_{\mathrm{F}} / L_0$ and can thereby be used to measure the length $L_0$ of
the resonator.

As visible in Fig.~\ref{fig:fabry-perot}, however, the amplitude of these
oscillations is strongly reduced as soon as the effective contact length
exceeds the length of one unit cell. One could view this situation as
\textit {smooth contacts} that cause the Fabry-P\'erot oscillations to be
broadened and the resonator length $L_0$ to be ill defined.

In some experiments using extended contacts on CNTs, the length of the
scattering regions was measured to be just as long as the uncovered region of
the tube,\cite{mann-btimnwrpoc2003} which could be explained based on our
model by a strong contact $\Delta$ and, therefore, a short effective contact
length. For weaker contacts $\Delta$, it is to be expected that Fabry-P\'erot
oscillations cannot be cleanly observed any more. A point defect inside the
contacted region might, of course, act as a scattering point instead and give
rise to oscillations that indicate a resonator longer than the uncovered
region of the CNT.

In our previous study,\cite{nemec-cdociicnaais2006} we chose to average
this oscillating conductance over $E_{\mathrm{F}} \pm 0.5 \tmop{eV}$ in order
to separate the finite-length Fabry-P\'erot effects from the effects caused by
the contacts themselves. Physically, this is similar to the thermal effects
caused by high enough temperature. For the chosen conductor length of $L_0 =
100 ~ \tmop{nm}$, this approach was very successful in canceling all
Fabry-P\'erot oscillations and in reproducing the physics of a single extended
contact. The resonance oscillations within the contact were, of course, also
strongly suppressed by the averaging, leaving only a minimal signature that we
correctly identified as such.

\tmfloatsmall{\resizebox{3.375in}{!}{\includegraphics{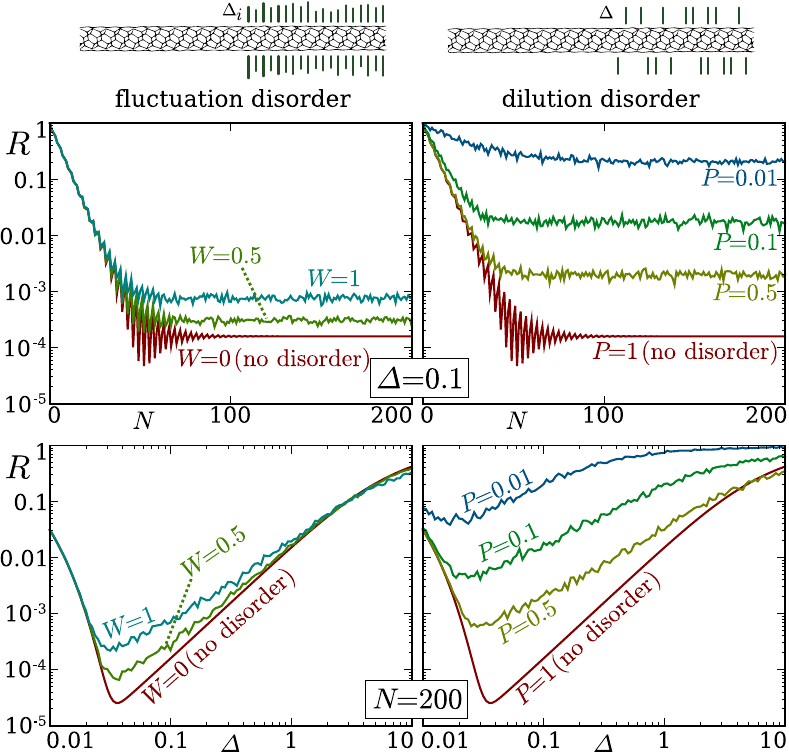}}}{\label{fig:non-epitaxial}Effects
of two different kinds of disorder on extended contacts: relatively weak
\textit {fluctuation disorder} with varying contact strength $\Delta_i$ on
each atom $i$ and stronger \textit {dilution disorder} with only a randomly
selected fraction of the atoms in the contact region attached to a lead. In
each case, the parameter $\Delta$ refers to the \textit {average} contact
strength. For details, see the text.}

\section{Nonepitaxial contacts}

Unlike the theoretical model contacts presented so far, realistic samples
produced in experiment are never perfectly epitaxial but contain imperfections
due to fabrication faults, lattice mismatch or metal faceting. To check
whether the effects described so far are robust to such perturbations, we have
investigated various kinds of disorder at the contact. A relatively weak
disorder was implemented as random fluctuations of the contact parameter
$\Delta$ on each atom $i$ as
\begin{eqnarray*}
  \Delta_i^{\tmop{fluct}} & = & \Delta \left( 1 + \xi_i^{\tmop{fluct}} W
  \right),
\end{eqnarray*}
with an evenly distributed random variable $- 1 \leqslant \xi_i^{\tmop{fluct}}
\leqslant 1$ and a parameter $W$ specifying the relative strength of the
fluctuations. As can be seen in Fig.~\ref{fig:non-epitaxial}, even for the
strongest possible value $W = 1$, the effect of the disorder is moderate and
purely quantitative.

An even stronger disorder was realized by using a model of diluted contacts,
where only a randomly selected fraction of the atoms in the contact region is
contacted,
\begin{eqnarray*}
  \Delta^{\tmop{diluted}}_i & = & \left\{ \begin{array}{cc}
    \Delta / P & \text{with probability $P$}\\
    0 & \text{with probability $1 - P$.}
  \end{array} \right.
\end{eqnarray*}
This kind of disorder modifies to the observed behavior to a much larger
degree, but even in the extreme case of a 1\% dilution (i.e. $P = 0.01$), the
general trend of the original model is well preserved (see
Fig.~\ref{fig:non-epitaxial}).

\section{Material related calculations}

To link the model results obtained so far to the physical properties of real
contact materials, we performed density functional theory (DFT) calculations
of Ti and Pd monolayers interacting with a graphene layer as described
before.\cite{nemec-cdociicnaais2006} We described the valence electrons by
Troullier-Martins pseudopotentials and used the Perdew-Zunger form of the
exchange-correlation functional in the local density approximation to DFT, as
implemented in the SIESTA code.\cite{soler-tsmfaioms2002} With a
double-zeta basis and a 100~Ry energy cutoff in the plane-wave expansions of
the electron density and potential, we found the total energy to be converged
to $\lesssim 1 \text{meV} / \text{atom}$. We performed a full structure
optimization to determine the equilibrium adsorption geometry, the adsorption
energy, and the local charge redistribution caused by the metal-graphene
interaction. Since the interatomic distances in bulk Pd (2.7~{\AA}) and Ti
(2.95~{\AA}) lie close to the honeycomb spacing in graphene (2.46~{\AA}), we
considered only epitaxial adsorption. For both Pd and Ti, we found a slight
preference for the sixfold hollow site on graphite. For Pd, we found the
equilibrium interlayer distance to be 3.2~{\AA}, consistent with a relatively
weak, mostly covalent bond energy of 0.3~eV per Pd atom. The interaction
between an epitaxial Ti monolayer and graphene was only insignificantly
stronger with 0.4~eV per Ti atom at an interlayer distance of 3.0~{\AA}.

\tmfloatsmall{\resizebox{3.375in}{!}{\includegraphics{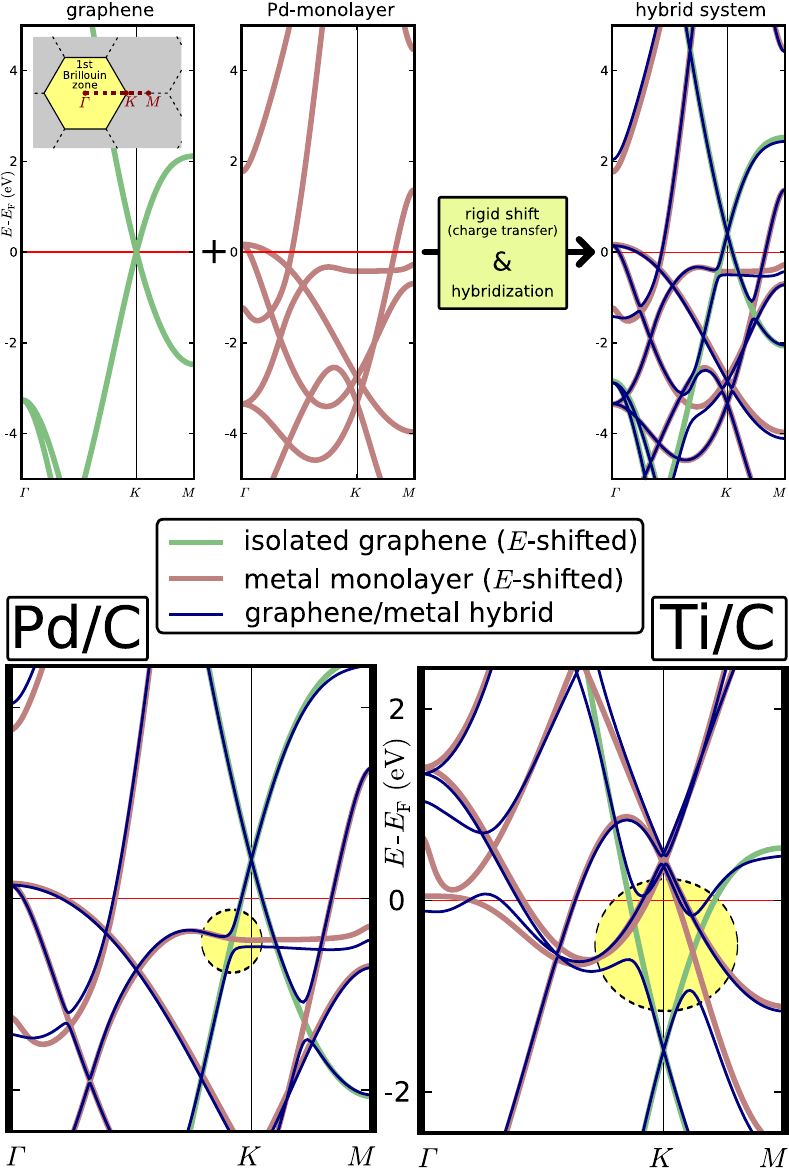}}}{\label{fig:PdC-TiC-bands}Analysis
of the hybridization between a graphene sheet and a metal monolayer. As
visible in the upper scheme, the hybrid band structure matches well with an
overlay of the band structures of the two individual systems rigidly shifted
in energy and hybridization at some band crossings. Highlighted in the Pd/C
and Ti/C band structures are the regions of interest, i.e., those
hybridizations that contribute most to the electron injection.}

To study the electronic coupling between the two systems, we first inspected
the band structure (see Fig.~\ref{fig:PdC-TiC-bands}). Especially for Pd as a
contacting metal, the extraction of parameters for our model is greatly
simplified by the fact that the band structure of hybrid lies close to the
superposition of the metal and carbon band structures. One can see a rigid
shift of the carbon bands by $E_{\mathrm{C}} = 0.374 ~ \text{eV}$ while the
palladium bands are shifted slightly in the opposite direction with $\Delta
E_{\tmop{Pd}} = - 0.020 ~ \text{eV}$. On top of this rigid shift, one can
observe slight hybridization effects in the band structure. For injecting
conduction electrons into a graphene sheet or the wall of a carbon nanotube,
the most important area of the Brillouin zone is the $K$ point, the Fermi
point of graphene. In Fig.~\ref{fig:PdC-TiC-bands}, a small avoided crossing
is visible near this region in the Pd/C band structure. To extract an estimate
of tight-binding parameters from these data, we modeled a honeycomb lattice
and a matching hexagonal lattice representing both sheets. As it turned out, a
single orbital per atom is sufficient to obtain bands that can be fitted to
the hybridizing bands near the Fermi level with a single parameter each. Now,
an additional coupling between the two sheets was introduced, linking each Pd
atom with its six neighboring C atoms. This hopping parameter could then be
tuned to reproduce a hybridization between the two subsystems, which is close
to that in the hybrid bandstructure obtained from DFT, resulting in a coupling
of $t_{\text{Pd/C}} \approx 0.15 ~ \text{eV}$.

For the case of Ti, the distortions in the band structure caused by the
hybridization of the two layers are considerably stronger than for Pd. Still,
the change in the carbon related levels can be modeled by a rigid shift of
$\Delta E_{\mathrm{C}} = - 1.15 ~ \tmop{eV}$. To determine the hopping
parameter, the same procedure as for Pd could not be directly applied, because
the relevant band of the Ti-monolayer cannot be reproduced with a
single-orbital hexagonal lattice. Instead, a rough estimate was obtained by
visually comparing the bandstructures themselves where the avoided crossing
near the $K$-point is at least twice as large as for Pd, giving an estimated
value of $t_{\text{Ti/C}} \gtrsim 0.3 ~ \text{eV}$.

To turn these parameters into values of $\Delta$ that can be directly placed
into our model calculations, we need the surface density of states, which is
comparable for both materials at $\mathcal{N}_{\text{Me}} \approx 1 ~
\text{eV}^{- 1}$. Finally, the connectivity at the interface is also
important: Each C atom contacted to three different metal atoms simply triples
the value of $\Delta$. The internal connections inside the metal are already
taken into account with the surface density of states and do not have to be
considered any further. With the relation $\Delta = t^2 \mathcal{N}$, this
again gives rough estimates of $\Delta_{\tmop{Pd}} \approx 0.06 ~ \text{eV}$
and $\Delta_{\tmop{Ti}} \gtrsim 0.3 ~ \tmop{eV}$.

Unfortunately, this approach of computing a graphene layer and a layer of the
contacting material within a common unit cell cannot necessarily be
transferred to other materials of interest in any straightforward way. There
exist, however, \textit {ab initio} calculations of various metals in contact
with graphene or CNTs that show a clear
trend:\cite{durgun-ssoaosaoacn2003,maiti-mibeawp2004,durgun-eaesoiaaocn2004}
The highly conducting metals Au, Ag and Cu generally have a very weak binding
energy, insufficient for wetting the carbon surface; so, a clean contact is
hard to achieve. Pd, Pt and Ti all have sufficient binding energies for
wetting the surface. Pt and Ti both have higher binding energies than Pd. For
pointlike end contacts, such strong bonds give good
transparency.\cite{liu-aisotscn2003} For extended contacts, however, the
weak bonds of Pd are to be preferred.

Previous calculations\cite{maiti-mibeawp2004} attributed the difference
between Ti and Pd metal contacts to the formation of different-sized metal
clusters at the interface. Our results, presented above, offer a more
fundamental explanation: It is exactly the weak bonding between Pd and
graphene or CNTs{\emdash}just large enough to wet the surface{\emdash}that
makes Pd such an excellent contact material.

\section{Conclusions}

To conclude, we introduced a model for electrical contacts for carbon
nanotubes and graphene nanoribbons that captures the fact that the contacts in
an experimental setup typically extend over a length of several tens of
nanometers, covering the carbon structure with some contact metal. We have
demonstrated the counterintuitive result that, given a metal coating of
several nanometer length, the contact transparency is actually improved by
using a metal that couples more weakly to the carbon surface. Using
\textit {ab initio} results of Ti and Pd as contact metals, we have
demonstrated that Pd actually forms a weaker bond, giving an explanation of
the experimental finding that Pd forms good contacts. This finding suggests a
possible route for future attempts in optimizing the charge injection in
carbon nanotubes and graphene, that is, to find contacting materials that
couple to the carbon surface as softly as possible to exploit the available
contact length.

Starting with a detailed analysis of an analytically solvable minimal model,
we have demonstrated by numerical calculations that the qualitative results
are robust to various modifications of the system. Replacing the atomic wire
by the actual atomic structure of a carbon nanotube or a graphene ribbon can
be accomodated by adjusting just two parameters at any given energy. Including
a second contact to model a realistic two-terminal conductance measurement
leads to Fabry-P\'erot oscillations that can be averaged out to give the
original result. A weak disorder in the contact interface has very little
quantitative effect, and even a strong disorder (similar to metal grains
forming contact only in certain points) leaves the qualitative behavior
unchanged.

In view of a possible experimental confirmation of our results, the biggest
challenge may lie in the fabrication of well-controlled finite-length contacts
down below the magnitude of the effective contact length of a few nanometers.
With current technology, this precision is yet out of reach, but with future
developments, it should well be possible to tune the geometry of contact with
sufficient precision. An alternative approach would be to tune the coupling
contact interface by some means. Direct tuning of the bonds, as it can be done
for molecular junctions,\cite{grter-rttacmjiale2005} seems unfeasible for
this kind of geometry. Instead, the insertion of an insulating atomic layer
below the contacting metal is already being used to improve the contacts for
graphene monolayers and might work as well for nanotubes.

\section*{Acknowledgements}

We acknowledge fruitful discussions with Ferdinand Evers, Sybille Gemming and
Christian Sch\"onenberger. This work was funded by the Volkswagen Foundation
under Grant No. I/78~340 and by the European Union grant CARDEQ under Contract
No. IST-021285-2. Support from the Vielberth Foundation is also gratefully
acknowledged. D.~T. acknowledges financial support by NSF~NIRT Grant No.
ECS-0506309, NSF~NSEC Grant No. EEC-425826, and the Alexander von Humboldt
Foundation.

\appendix\section{Analytics for the one-dimensional
model\label{full-solution}}

\subsection{Transmission calculations}

The Hamiltonian of a two-probe system for transport calculations is given by
\begin{eqnarray}
  \mathcal{H} & = & \left( \begin{array}{ccc}
    H_{\mathrm{L}} & H_{\mathrm{\tmop{Lc}}} & 0\\
    H_{\mathrm{\tmop{cL}}} & H_{\mathrm{c}} & H_{\mathrm{\tmop{cR}}}\\
    0 & H_{\mathrm{\tmop{Rc}}} & H_{\mathrm{R}}
  \end{array} \right)  \label{eqn:hamiltonian-twoprobe}
\end{eqnarray}
where $H_{\mathrm{} \mathrm{c}}$ describes the finite-size conductor region
and $H_{\mathrm{L} / \mathrm{R}}$ describes the leads, which are connected to
independent reservoirs and have no direct contact with each other. From
$\mathcal{H}=\mathcal{H}^{\dag}$ it follows that
$H_{\mathrm{\tmop{Lc}}}^{\phantom{\dag}} = H_{\mathrm{\tmop{cL}}}^{\dag}$ and
$H_{\mathrm{\tmop{Rc}}}^{\phantom{\dag}} = H_{\mathrm{\tmop{cR}}}^{\dag}$.

To simplify the notation, we first define the \textit {complex-energy Green
function}:
\begin{eqnarray*}
  \mathcal{G} \left( \mathcal{E} \right) & = & \left( \mathcal{E}-\mathcal{H}
  \right)^{- 1}
\end{eqnarray*}
and derive from it the expressions for the retarded and advanced Green
functions ($\mathcal{E}= E \pm \mathi \eta$)
\begin{eqnarray*}
  \mathcal{G}^{\mathrm{r}} \left( E \right) & = & \lim_{\eta \rightarrow 0^+}
  \mathcal{G} \left( E + \mathi \eta \right),\\
  \mathcal{G}^{\mathrm{a}} \left( E \right) & = & \lim_{\eta \rightarrow 0^+}
  \mathcal{G} \left( E - \mathi \eta \right) .
\end{eqnarray*}
The \textit {transmission} through this system is given
by\cite{fisher-rbcatm1981,datta-etims1999}
\begin{eqnarray}
  T & = & \tmop{Tr} \left\{ \Gamma_{\mathrm{L}}
  \mathcal{G}_{\mathrm{c}}^{\mathrm{r}} \Gamma_{\mathrm{R}}
  \mathcal{G}_{\mathrm{c}}^{\mathrm{a}} \right\},  \label{fisher-lee}
\end{eqnarray}
with
\begin{eqnarray*}
  \mathcal{G}_{\mathrm{c}} & = & \left( \mathcal{E}- H_c - \Sigma_{\mathrm{L}}
  - \Sigma_{\mathrm{R}} \right)^{- 1},\\
  \Sigma_{\alpha} & = & H_{\mathrm{c}, \alpha} \mathcal{G}_{\alpha} H_{\alpha,
  \mathrm{c}}, \hspace{1.5cm} \alpha = \mathrm{L}, \mathrm{R},\\
  \Gamma_{\alpha} & = & \mathi \left( \Sigma_{\alpha}^{\mathrm{r}} -
  \Sigma_{\alpha}^{\mathrm{a}} \right),\\
  \mathcal{G}_{\alpha} & = & \left( \mathcal{E}- H_{\alpha} \right)^{- 1} .
\end{eqnarray*}
The Hamiltonian of the model at hand, depicted in
Fig~\ref{fig:linchain-scheme}, can be split up according to
Eq.~(\ref{eqn:hamiltonian-twoprobe}): The conductor consists of just one atom,
so its Hamiltonian is a $1 \times 1$ matrix $H_{\mathrm{c}} = \left(
\begin{array}{l}
  \varepsilon
\end{array} \right)$. The left lead is a semi-infinite chain, contacted only
at the last atom,
\begin{eqnarray*}
  H_{\mathrm{L}} & = & \left( \begin{array}{ccccc}
    & \ddots & \ddots & \ddots & 0\\
    \cdots & 0 & - \gamma & \varepsilon & - \gamma\\
    & \cdots & 0 & - \gamma & \varepsilon
  \end{array} \right)_{\infty \times \infty},\\
  H_{\mathrm{\tmop{cL}}} & = & \left( \begin{array}{lllll}
    \phantom{\ldots} & \cdots & \phantom{-} 0 & \phantom{-} 0 & - \gamma
  \end{array} \right)_{1 \times \infty} .
\end{eqnarray*}
The right lead consists of a chain of $N$ atoms, each attached to a wideband
lead. This can be captured by defining an \textit {effective Hamiltonian} of
the form
\begin{eqnarray}
  H_{\mathrm{R}}^{\mathrm{\tmop{eff}}} & = & \left( \begin{array}{ccccc}
    \varepsilon - \frac{\mathi \Delta}{2} & - \gamma &  & \cdots & 0\\
    - \gamma & \varepsilon - \frac{\mathi \Delta}{2} & \ddots &  & \vdots\\
    & - \gamma & \ddots & - \gamma & \\
    \vdots &  & \ddots & \varepsilon - \frac{\mathi \Delta}{2} & - \gamma\\
    0 & \cdots &  & - \gamma & \varepsilon - \frac{\mathi \Delta}{2}
  \end{array} \right)_{N \times N}  \label{H-R-eff}
\end{eqnarray}
together with a contact point in the first atom only,
\begin{eqnarray*}
  H_{\mathrm{\tmop{cR}}} & = & \left( \begin{array}{llll}
    - \gamma \phantom{\ldots} 0 & 0 & \cdots & 0
  \end{array} \right)_{1 \times N} .
\end{eqnarray*}
Note that $H_{\mathrm{R}}^{\mathrm{\tmop{eff}}}$ is the effective Hamiltonian
containing the \textit {retarded} self-energy, so
$\mathcal{G}^{\mathrm{r}}_{\mathrm{R}} = \left( E + \mathi 0^+ -
H_{\mathrm{R}}^{\mathrm{\tmop{eff}}} \right)^{- 1}$ and
$\mathcal{G}^{\mathrm{a}}_{\mathrm{R}} = \left( E - \mathi 0^+ -
(H_{\mathrm{R}}^{\mathrm{\tmop{eff}}})^{\dag} \right)^{- 1}$.

In the following, we will simplify the notation by setting $\gamma = 1$ and
$\varepsilon = 0$. Both constants can be reintroduced in the final result
[Eq.~(\ref{transmission-chain-to-CWBL})] by substituting $E \rightarrow \left(
E - \varepsilon \right) / \gamma$.

\subsection{Inverse based on Chebyshev polynomials}

As a starting point for a full analytical solution, we look at a finite linear
chain of length $N$, which has the Hamiltonian
\begin{eqnarray*}
  \mathcal{H}^N & = & \left( \begin{array}{ccccc}
    0 & - 1 &  & \cdots & 0\\
    - 1 & 0 & \ddots &  & \vdots\\
    & - 1 & \ddots & - 1 & \\
    \vdots &  & \ddots & 0 & - 1\\
    0 & \cdots &  & - 1 & 0
  \end{array} \right)_{N \times N} .
\end{eqnarray*}
The quantity of interest of this system is the $1, 1$ matrix element of the
Green function $\mathcal{G}_{}^N \left( \mathcal{E} \right) = \left(
\mathcal{E}-\mathcal{H}_N \right)^{- 1}$. The solution is based on the
Chebyshev polynomials of the second kind,\cite{gradshteyn-toisap2000} which
can be defined via the determinant identity,
\begin{eqnarray*}
  U_n \left( x \right) & = & \det \left( \begin{array}{ccccc}
    2 x & 1 &  &  & 0\\
    1 & 2 x & \ddots &  & \\
    & 1 & \ddots & 1 & \\
    &  & \ddots & 2 x & 1\\
    0 &  & \phantom{\ddots} & 1 & 2 x
  \end{array} \right)_{n \times n},
\end{eqnarray*}
or, equivalently, by the recursive definition,
\begin{eqnarray}
  U_0 \left( x \right) & = & 1, \nonumber\\
  U_1 \left( x \right) & = & 2 x, \nonumber\\
  U_{n + 1} \left( x \right) & = & 2 xU_n \left( x \right) - U_{n - 1} \left(
  x \right) .  \label{eqn:chebyshev-recursive}
\end{eqnarray}
We can now use the well-known identity for the matrix inverse,

\begin{widetext}
\begin{eqnarray*}
  \left( A^{- 1} \right)_{i j} & = & \left. \frac{1}{\det (A)} \det \left(
  \begin{array}{ccccccc}
    A_{1, 1} & \cdots & A_{1, j - 1} & 0 & A_{1, j + 1} & \cdots & A_{1, N}\\
    \vdots & \ddots & \vdots & \vdots & \vdots & \ddots & \vdots\\
    A_{i - 1, 1} & \cdots & A_{i - 1, j - 1} & 0 & A_{i - 1, j + 1} & \cdots &
    A_{i - 1, N}\\
    0 & \cdots & 0 & 1 & 0 & \cdots & 0\\
    A_{i + 1, 1} & \cdots & A_{i + 1, j - 1} & 0 & A_{i + 1, j + 1} & \cdots &
    A_{i + 1, N}\\
    \vdots & \ddots & \vdots & \vdots & \vdots & \ddots & \vdots\\
    A_{N, 1} & \cdots & A_{N, j - 1} & 0 & A_{N, j + 1} & \cdots & A_{N, N}
  \end{array} \right)_{} \right.
\end{eqnarray*}
\end{widetext}

to find
\begin{eqnarray*}
  \left[ \mathcal{G}^N \left( \mathcal{E} \right) \right]_{i, j} & = & \left[
  \left( \mathcal{E}-\mathcal{H}_N \right)^{- 1} \right]_{i, j}\\
  & = & \left[ \mathcal{G}^N \left( \mathcal{E} \right) \right]_{j, i}\\
  & \overset{i \leqslant j}{=} & \left( - 1 \right)^{i - j} \frac{U_{i - 1}
  \left( \mathcal{E}/ 2 \right) U_{N - j} \left( \mathcal{E}/ 2 \right)}{U_N
  \left( \mathcal{E}/ 2 \right)}
\end{eqnarray*}
and, specifically,
\begin{eqnarray}
  \left[ \mathcal{G}^N \left( \mathcal{E} \right) \right]_{1, 1} & = &
  \frac{U_{N - 1} \left( \mathcal{E}/ 2 \right)}{U_N \left( \mathcal{E}/ 2
  \right)} \nonumber\\
  & = : & f_N \left( \mathcal{E}/ 2 \right) .  \label{define-f_N}
\end{eqnarray}

\subsection{Surface of semi-infinite linear chain}

The surface Green function of a semi-infinite linear chain can be defined as
\begin{eqnarray*}
  \mathcal{G}_s \left( \mathcal{E} \right) & = & \lim_{N \rightarrow \infty}
  \left[ \mathcal{G}^N \left( \mathcal{E} \right) \right]_{1, 1}\\
  & = & \lim_{N \rightarrow \infty} f_N \left( \mathcal{E}/ 2 \right)\\
  & = : & f_{\infty} \left( \mathcal{E}/ 2 \right) .
\end{eqnarray*}
To find an expression for $f_{\infty} \left( x \right)$, we can use the
recursive definition of the Chebyshev polynomials
[Eq.~(\ref{eqn:chebyshev-recursive})] and obtain
\begin{eqnarray*}
  f_N \left( x \right) & = & \left( 2 x - f_{N - 1} \left( x \right)
  \right)^{- 1} .
\end{eqnarray*}
For $N \rightarrow \infty$, this becomes
\begin{eqnarray*}
  f_{\infty} \left( x \right) & = & \left( 2 x - f_{\infty} \left( x \right)
  \right)^{- 1},
\end{eqnarray*}
which has two solutions $f_{\infty} \left( x \right) = x \left( 1 \pm \sqrt{1
- 1 / x^2} \right)$. On the real axis, it follows from
Eqn.~(\ref{eqn:chebyshev-recursive}) by induction that $\left| f_N \left( x
\right) \right| < 1$ when $\left| x \right| > 1$, so that we can select the
correct solution
\begin{eqnarray}
  f_{\infty} \left( x \right) & = & x \left( 1 - \sqrt{1 - 1 / x^2} \right), 
  \label{f-infty}
\end{eqnarray}
which can be continued analytically to $x \in \mathbb{C}\backslash(- 1, 1)$
by reading the square root of a complex number as the \textit {principal
square root}, uniquely defined everywhere except on the negative real axis by
the condition $\tmop{Re} \left( \sqrt{x} \right) \geqslant 0, \hspace{10pt}
\forall x \in \mathbb{C}$.

The retarded surface Green function follows as
\begin{eqnarray}
  \mathcal{G}_s^r \left( E \right) & = & \lim_{\eta \rightarrow 0^+}
  \mathcal{G}_s \left( E + \mathi \eta \right) \nonumber\\
  & = & \left\{ \begin{array}{lll}
    E / 2 - \sqrt{E^2 / 4 - 1} & \text{for} & \left| E \right| \geqslant 2\\
    E / 2 - \mathi \sqrt{1 - E^2 / 4} & \text{for} & \left| E \right|
    \leqslant 2
  \end{array}  \label{G-s-r} \right.
\end{eqnarray}

\subsection{Transmission of the model system}

With these results, we can now obtain the quantum mechanical transmission of
our model system. The left lead is a semi-infinite chain giving a self
self-energy of
\begin{eqnarray*}
  \Sigma_{\mathrm{L}} & = & H_{\mathrm{\tmop{cL}}} \left( \mathcal{E}-
  H_{\mathrm{L}} \right)^{- 1} H_{\mathrm{\tmop{Lc}}}\\
  & = & \mathcal{G}_s \left( \mathcal{E} \right) .
\end{eqnarray*}
For $\left| E \right| \geqslant 2$, $\mathcal{G}_s^{\mathrm{r}} \left( E
\right)$ is real [see Eq. (\ref{G-s-r})], so, $\Gamma_{\mathrm{L}}$ and with
it by Eq. (\ref{fisher-lee}) the whole transmission $T$, are strictly zero. In
the following, we therefore assume $\left| E \right| < 2$ and select the
second case in Eq.~(\ref{G-s-r}):
\begin{eqnarray*}
  \Sigma_{\mathrm{L}}^{\mathrm{r}} & = & \frac{E}{2} \left( 1 - \mathi \sqrt{4
  / E^2 - 1} \right)
\end{eqnarray*}
To find the self-energy of the right lead, we use the definition of
$H_{\mathrm{R}}^{\mathrm{\tmop{eff}}}$ from Eq. (\ref{H-R-eff}) and find
\begin{eqnarray*}
  \Sigma_{\mathrm{R}}^{\mathrm{r}} & = & H_{\mathrm{\tmop{cR}}} \left( E - H^N
  + \mathi \Delta / 2 \right)^{- 1} H_{\mathrm{\tmop{Rc}}}\\
  & = & \left[ \mathcal{G}^N \left( E + \mathi \Delta / 2 \right) \right]_{1,
  1}\\
  & = & f_N \left( E / 2 + \mathi \Delta / 4 \right) .
\end{eqnarray*}
Now, we can put together all parts to calculate the transmission,
\begin{eqnarray}
  \mathcal{G}^{\mathrm{r}}_{\mathrm{c}} & = & \left( E -
  \Sigma_{\mathrm{L}}^{\mathrm{r}} - \Sigma_{\mathrm{R}}^{\mathrm{r}}
  \right)^{- 1} \nonumber\\
  & = & 2 \left( E + \mathi E \sqrt{4 / E^2 - 1} - 2 f_N \left( E / 2 +
  \mathi \Delta / 4 \right) \right)^{- 1}, \nonumber\\
  \mathcal{G}^{\mathrm{a}}_{\mathrm{c}} & = & 2 \left( E - \mathi E \sqrt{4 /
  E^2 - 1} - 2 f_N \left( E / 2 - \mathi \Delta / 4 \right) \right)^{- 1},
  \nonumber\\
  \Gamma_L & = & \mathi \left( \Sigma_{\mathrm{L}}^{\mathrm{r}} -
  \Sigma_{\mathrm{L}}^{\mathrm{a}} \right) \nonumber\\
  & = & E \sqrt{4 / E^2 - 1}, \nonumber\\
  \Gamma_R & = & \mathi \left[ f_N \left( E / 2 + \mathi \Delta / 4 \right) -
  f_N \left( E / 2 - \mathi \Delta / 4 \right) \right] \nonumber\\
  & = & 2 \tmop{Im} \left[ f_N \left( E / 2 - \mathi \Delta / 4 \right)
  \right] \nonumber\\
  T & = & \frac{8 \sqrt{4 - E^2} \tmop{Im} \left( f_N \left( E / 2 - \mathi
  \Delta / 4 \right) \right)}{\left| E - \mathi \sqrt{4 - E^2} - 2 f_N \left(
  E / 2 - \mathi \Delta / 4 \right) \right|^2} 
  \label{transmission-chain-to-CWBL}
\end{eqnarray}


\end{document}